\documentclass[aps,pra,twocolumn,groupedaddress]{revtex4-1}

\usepackage{graphicx}
\usepackage{color}
\usepackage{dcolumn}
\usepackage{amsmath}
\usepackage{mathtools}
\usepackage{amssymb}
\usepackage{amsfonts}
\usepackage{bm}
\usepackage{epstopdf}
\usepackage{picture}
\usepackage{enumitem}
\usepackage{afterpage}
\usepackage{placeins}
\usepackage{float}
\usepackage{caption}
\usepackage{cases}
\usepackage{xr}
\usepackage[normalem]{ulem}

\newcommand{\Fvec}{\mathbf{F}}
\newcommand{\fvec}{\mathbf{f}}

\newcommand{\Kmat}{\mathbf{K}}
\newcommand{\kmat}{\mathbf{k}}

\newcommand{\vvec}{\mathbf{v}}
\newcommand{\xvec}{\mathbf{x}}

\newcommand{\Xvec}{\mathbf{X}}

\begin{document}

\title{PT-like phase transition and limit cycle oscillations in non-reciprocally coupled optomechanical oscillators levitated in vacuum}

\author{Vojtěch Liška$^1$}  
\author{Tereza Zemánková$^1$}  
\author{Petr J\'{a}kl$^1$} 
\author{Martin \v{S}iler$^1$}
\author{Stephen H. Simpson$^1$}
\email{simpson@isibrno.cz}
\author{Pavel Zem\'anek$^1$}
\author{Oto Brzobohat\'y$^1$}
\email{otobrzo@isibrno.cz}
\affiliation{$^1$The Czech Academy of Sciences, Institute of Scientific Instruments, Kr\'{a}lovopolsk\'{a} 147, 612 64 Brno, Czech Republic}

\begin{abstract}

Nanoparticles levitated in an optical trap provide a versatile platform to study mechanical oscillators in a controlled environment with tuneable parameters. Recently, it has become possible to couple two of these optomechanical oscillators. Here, we demonstrate the collective non-Hermitian dynamics of such a pair of non-conservatively coupled oscillators. We take advantage of the tunability of the optical interactions between the particles in our system and set the optical interaction between the particles to be purely non-reciprocal. By continuously varying the relative power of the trapping beams, we take the system through a transition, similar to a parity-time phase transition. A Hopf bifurcation at a critical point results in the formation of collective limit cycle oscillations, resembling those observed in phonon lasers. These coupled levitated oscillators provide a platform for exceptional point optomechanical sensing and can be extended to multi-particle systems, paving the way for the development of topological optomechanical media.
 
\end{abstract}

\maketitle

\section{Introduction}\label{sec1}

The ability of focused laser beams to confine, manipulate, and control the motion of mesoscopic particles under vacuum conditions has turned the field of levitational optomechanics into a powerful tool for addressing crucial questions in the physical sciences, ranging from the macroscopic limits of quantum mechanics to the thermodynamic limits of computation \cite{millen_quantum_2020,gonzalez-ballestero_levitodynamics_2021}. Of particular significance are the recent achievements of ground state cooling of single \cite{delic_cooling_2020} and multiple \cite{piotrowski_simultaneous_2023,pontin_simultaneous_2023} degrees of freedom of isolated particles. These experiments exploit the potential-like, conservative properties of optical forces, which ensure a base level of dynamic and thermodynamic stability. 

However, since light is a flow of momentum, optical forces are intrinsically non-conservative \cite{sukhov_non-conservative_2017}. Recent work emphasises this characteristic \cite{li_non-hermitian_2021}, which appears whenever simple symmetries are broken. Examples include isotropic particles in circularly polarized beams \cite{svak_transverse_2018}, birefringent or non-spherical particles in linearly or circularly polarized beams \cite{arita_coherent_2020,arita_cooling_2023, hu_structured_2023}, and optically interacting particles in beams with phase decoherence \cite{rieser_tunable_2022}. The forces in these systems can be locally described by a generalized Hooke's law, having a non-symmetric stiffness matrix, resulting in biased stochastic motion \cite{simpson_first-order_2010}. For sufficiently high driving, or low dissipation, this bias grows until inertial forces overcome attractive forces causing a bifurcation, the formation of a limit cycle oscillation \cite{simpson_stochastic_2021} and, in multi-particle systems, synchronization \cite{brzobohaty_synchronization_2023}.

These effects can be conveniently situated within the framework of non-Hermitian physics \cite{ashida_non-hermitian_2020,el-ganainy_non-hermitian_2018, feng_non-hermitian_2017, ozturk_observation_2021, miri_exceptional_2019}, which is used to describe open systems (that exchange energy with their environments) in both the quantum and classical domains. In the latter case, complex photonic systems have received great attention \cite{feng_non-hermitian_2017}, with steadily growing interest in mechanical materials \cite{huber_topological_2016,mao_maxwell_2018,brandenbourger_non-reciprocal_2019}. 
Of particular interest are non-Hermitian systems with PT symmetry, i.e. those with Hamiltonians that commute with the parity-time ($\hat P \hat T$) operator. 
Here, we are mainly concerned with a key feature common to PT symmetric systems, the PT phase transition, where a system goes from an equilibrium state, described by eigenvalues that are real and distinct, to a non-equilibrium state for which the eigenvalues are no longer purely real. 

In many-body systems non-Hermitian properties, exemplified by those outlined above, underpin non-trivial topological effects such as the non-Hermitian skin effect (NHSE) \cite{zhang_review_2022}  or time crystallinity \cite{heugel_classical_2019,sacha_time_2017, wilczek_quantum_2012, shapere_classical_2012}. Particularly relevant here, is the recent work of Zheludev, demonstrating discrete photonic time crystals \cite{liu_photonic_2023}, the underlying mechanism for which is rooted in non-reciprocal forces similar to those discussed here \cite{raskatla_continuous_2023}.

In this article, we explore the stochastic dynamics of a pair of optomechanical oscillators with non-reciprocal coupling, optically levitated in vacuum. The system exhibits a trivial form of PT symmetry, slightly modified by viscous drag. Within the broken PT-symmetric phase, the eigenvalues describing the motion form a complex conjugate pair, resulting in an instability and limit cycle formation for sufficiently weak dissipation.

Limit cycle oscillations are the simplest form of regular dynamic attractor, consisting of closed loops in phase space (here, a four dimensional space consisting of the coordinates and velocities or momenta of the particles), onto which neighbouring trajectories converge. First identified by Poincare in the nineteenth century, they are ubiquitous in nature, appearing in diverse contexts ranging from the dynamics of neurons \cite{wilson_excitatory_1972} to astrophysical accretion discs \cite{shen_evolution_2014}. Significantly, the limit cycles formed in our system constitute a coherent excited state combining the motion of both particles. As we describe, this is readily extendable coherent states involving the motion of many particles, allowing future access to topological many body effects (see  Supplementary note 10).
We probe its changes in the stochastic dynamics with varying parameters, revealing transitional behaviour, that features the characteristic coalescence of the eigenvalues which describe the motion.

\section{Results}\label{sec2}

\begin{figure*}[htb]%
  \centering
  \includegraphics[width=\textwidth]{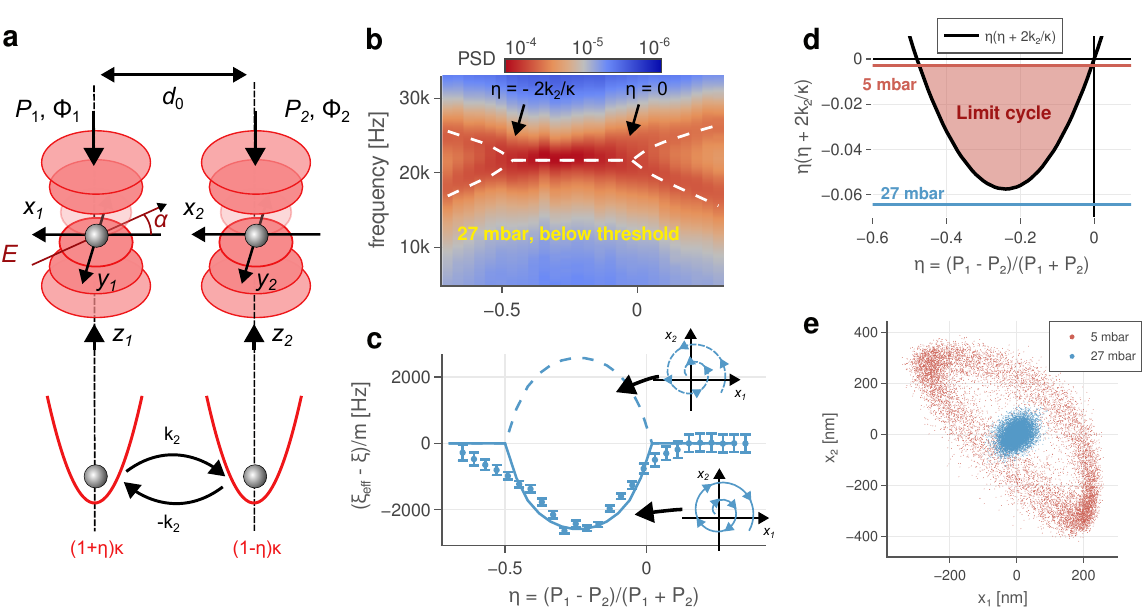}
  \caption{{\bf Overview of the experiment.} {\bf a,} Two silica nanoparticles oscillate in two independent standing wave optical traps. The interaction between both nanoparticles is mediated mainly by the light scattered between them. We set the conditions to make the interaction dominantly non-reciprocal.
  {\bf b,} The oscillation frequency of the $i^{\mathrm{th}}$  nanoparticle depends primarily on the power $P_i$ of the corresponding trapping laser and their frequency detuning is experimentally controlled by the power detuning $\eta=(P_1-P_2)/(P_1+P_2)$.  The sum of the power spectral densities (PSDs) of nanoparticles $x$-positions is encoded in colour. The white dashed curve corresponds to real value of eigenfrequency $\Re(\omega_i)=\sqrt{P_\mathrm{tot}/2m} \Re(\sqrt{\lambda_{1,2}})/2\pi$ from Eq. (\ref{eq:omi1}). 
  The degeneracy region, where both nanoparticles oscillate with the same frequency, can be found between two exceptional points corresponding to $\eta_1 = 0$ and $\eta_2 = -2k_2/\kappa$ from Eq. (\ref{eq:lambda}). 
  {\bf c,} The effective damping of quasi modes $\xi_\mathrm{eff} \propto \Im(\omega_i)$ are non-degenerated in between the exceptional points. The dashed curve illustrates increased damping  where non-conservative force works against the motion of particles (spiraling into fixed point) and the full curve illustrates driving ($\Im(\omega_i)<0$) of the particles (spiraling outwards, i.e. the Hopf bifurcation), see Eq. (\ref{eq:omi1}). The blue circles denotes the value of effective damping determined from particles trajectories (the damped mode was not determined). These mean values together with standard deviation (the errobars) were determined using least squares method.
  {\bf  d,} The red-coloured region illustrates the theoretically predicted condition for the Hopf bifurcation beyond which the collective limit cycle emerges. For the lower pressure of 5\,mbar the limit cycle emerges for a wide range of power detunings $\eta$. However, for the larger pressure of 27\,mbar, the condition $\eta(\eta+2k_2/\kappa)<-\xi^2/(mP_\mathrm{tot}\kappa)$ is not fulfilled and both particles are in fixed point and the limit cycle is not developed. 
  {\bf e,}  Red dots illustrate the collective limit cycle using $x_{1,2}$ coordinates plotted for $\eta = -0.25$ while the blue dots denote the particle motion near the fixed point.}
  \label{fig1}
\end{figure*}

Our experimental system consists of a parallel pair of counter-propagating (CP) linearly polarized Gaussian optical beams with wavelength $\lambda=1550$\,nm, and beam waist radius, $w_0=1.5\,\mu$m, separated by a distance, $d_0$ in the $x$ direction. The total trapping power $P_\mathrm{tot} = P_1 + P_2 = 140$\,mW is divided into two independent optical traps in a vacuum chamber, which allows us to independently set the trapping stiffness of both traps $\kappa_i$, see Fig.~\ref{fig1}a and Methods. Each counter-propagating beam consists of stacks of interference fringes oriented normally to the beam axes with axial separation $\Delta z=\lambda/2$, and with transverse circular cross-sections, within which the optical intensity has a Gaussian profile. A silica nanosphere with radius, $a=305$\,nm, is confined by optical gradient forces within the middle fringe of each CP beam. Due to the shape of the local intensity distribution, the gradient of the force in the axial $z$ direction is much higher than that in transverse directions \cite{liska_cold_2023}. When the polarization direction is normal to the beam separation $\alpha = \pi/2$ (i.e. parallel to the $y$ direction), optical interactions between the spheres are maximized \cite{rieser_tunable_2022,liska_cold_2023}, see Fig~\ref{fig1}a. The dynamical effects of interest are most conspicuous in the direction in which the mechanical susceptibility is greatest. That is, the vibrational amplitudes are greatest in the $x$ direction, with non-equilibrium motion in the $z$ direction \cite{rieser_tunable_2022} also being present, but with an amplitude suppressed by the higher stiffness. The stochastic motion is qualitatively determined by the linearized Langevin equation, which in time domain is

\begin{equation}\label{eq:Langt0}
m \ddot \xvec(t)=-\Kmat \xvec(t) - \xi \dot \xvec(t) + \fvec^L(t),  
\end{equation}
and its Fourier space image is
\begin{equation}
-m \omega^2 \Xvec(\omega) = -\Kmat \Xvec(\omega) - \mathrm{i} \omega \xi \Xvec(\omega) + \Fvec^L(\omega) \label{eq:Langf0},
\end{equation}
where $\xvec=(x_1,x_2)$ and $\Xvec=(X_1,X_2)$  are the $x$ coordinates of the particles in the time and frequency domains respectively and $\fvec^L$ and $\Fvec^L$ are the corresponding noise terms, $\Kmat$ is a stiffness matrix representing the linearised force \cite{svak_transverse_2018,arita_coherent_2020} and $\xi$ is the Stokes drag, proportional to pressure in the regime of interest. A thorough analysis of the general form of the stiffness matrix, $\Kmat$, in both the dipole approximation and Mie regimes, describing conservative and non-conservative contributions, and their variation with optical power distribution and overall trap geometry is provided in Supplementary Note 2. 

For the current purposes, we adjust the spacing between the traps to suppress the conservative contribution to the stiffness (in the experiment this corresponds to a default value of $d_0=8.6\,\mu$m) 
 and set the relative phase of the beams, $\Delta \Phi$, to $\pi/2$, maximising the non-conservative coupling \cite{rieser_tunable_2022}. This leaves a purely non-reciprocal interaction (see Fig. \ref{fig1}a bottom), 
\begin{equation}\label{eq:KK}
    \Kmat \approx \frac{1}{2} P_\mathrm{tot} 
    \begin{bmatrix}
        (1+\eta)\kappa + k_2 & -k_2 \\ k_2 & (1-\eta)\kappa - k_2 
    \end{bmatrix}.
\end{equation}

\begin{figure*}[htb]%
  \centering
  \includegraphics[width=\textwidth]{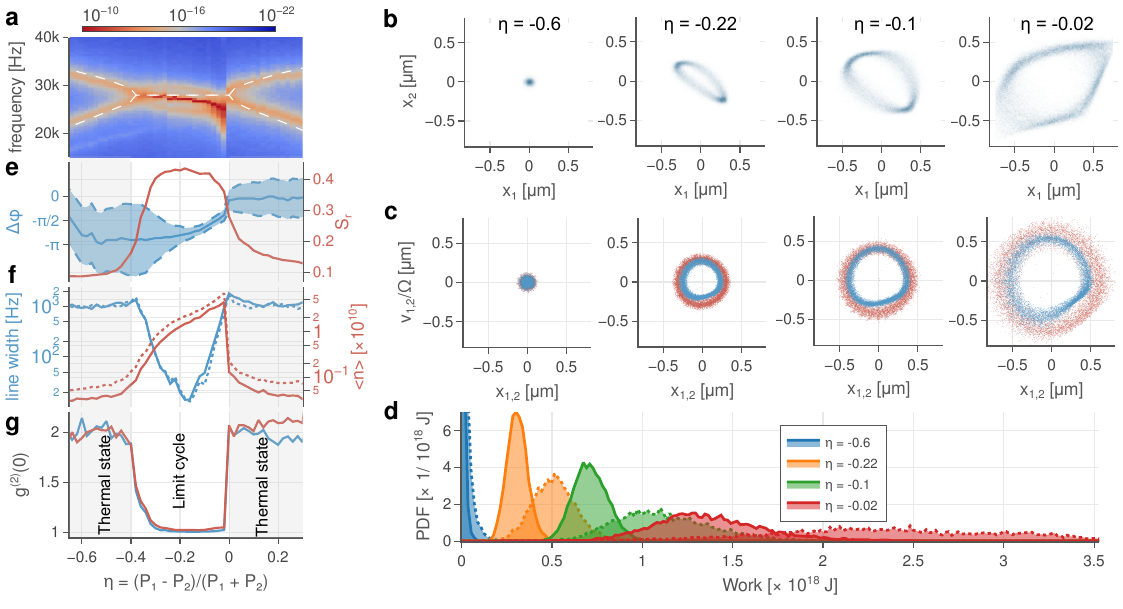} 
  \caption{{\bf Above threshold behaviour at the pressure of 5\,mbar.} 
  {\bf a,}\,The sum of the power spectral densities of nanoparticles $x$-positions (PSD), where the limit cycle emerged, for different power detuning $\eta$. 
  {\bf b,} and {\bf c,} Demonstration of the emerging collective limit cycle as the $x_1$, $x_2$ plot and the projection of phase space trajectories of individual particles (red and blue colour) for different $\eta$ close or inside the degeneracy region.  
  {\bf c,}\,The phase space trajectories of individual particles projected into a plane of velocity and position (blue and red for particles 1 and 2, respectively) illustrate the transition from a thermal state ($\eta=-0.6$) to coherent states for the same $\eta$ as in {\bf b}. 
  {\bf d,}\,Probability density function of the work done by the individual particle (full curve for particle 1 and dotted curve for particle 2) during one limit cycle period determined for the same $\eta$ as in {\bf b}. The changes of the distributions illustrate the transition from a thermal state ($\eta=-0.6$) to coherent states.  
  A non-zero value of the work corresponds to the pumping of energy to the system from the laser beams and its dissipation via drag force.  
  {\bf e,}\,The mean value (full blue line) with standard deviation (blue dashed lines) of the phase difference between particles motion $\Delta \varphi$ together with the value of the relative Shannon entropy $S_r$ (red) illustrating the phase locking of the collective motion of the particles. 
  {\bf f,}\,The width of the PSD peaks in {\bf a} significantly decreases from $10^3$ Hz to 10 Hz in the degeneracy region together with the rapid increase of phonon population $\langle n \rangle \propto \langle x_i^2 \rangle$. Full and dotted curves correspond to the values for the nanoparticle 1 and 2, respectively.  
  {\bf g,}\,The second-order phonon auto-correlation at zero time delay $g^{(2)}(0)$ illustrating the transition from the thermal state ($g^{(2)}(0)=2$) to the coherent  state of the limit cycle ($g^{(2)}(0)=1$).}\label{fig2}
\end{figure*} 
Here, the power detuning is $\eta=(P_1-P_2)/(P_1+P_2)=(P_1-P_2)/P_\mathrm{tot}$, where $P_{1,2}$ is the optical power in traps 1 and 2, $k_2 \propto \sin(kd_0)\sin(\Delta \Phi)$ is the non-reciprocal coupling rate and $\Delta \Phi=(\Phi_1-\Phi_2)$ is the relative optical phase of the beams. \\
If the weak viscous drag is neglected, along with the accompanying thermal fluctuations, then the linearized equations of motion, Eq. (\ref{eq:Langt0}) are trivially PT symmetric. This is the underlying reason for the phenomena we observed. Including viscous drag terms breaks this symmetry. However, the balance of gain and loss processes, a key feature of PT symmetric systems, is retained and the system continues to exhibit characteristic behaviour. These points are discussed more fully in Supplementary Note 7.\\
Under the condition, $\xi=0$, the oscillation eigen-frequencies in Eq. (\ref{eq:Langf0}), are determined by the purely real, secular equation, $|\Kmat-m\omega^2|=0$. The corresponding oscillation eigen-frequencies are,
\begin{equation}\label{eq:omi0}
\omega_i=\pm \sqrt{\frac{P_\mathrm{tot}}{2m}\lambda_{1,2}},
\end{equation}
where $i=1,\dots, 4$, and $\lambda_{1,2}$ are the eigenvalues of the power normalized stiffness, $\kmat=2\Kmat / P_\mathrm{tot}$, 
\begin{equation}\label{eq:lambda}
\lambda_{1,2}=\kappa \pm \kappa \sqrt{\eta\big(\eta+2k_2/\kappa \big)}.
\end{equation}
\begin{figure*}[htb]%
  \centering
  \includegraphics[width=\textwidth]{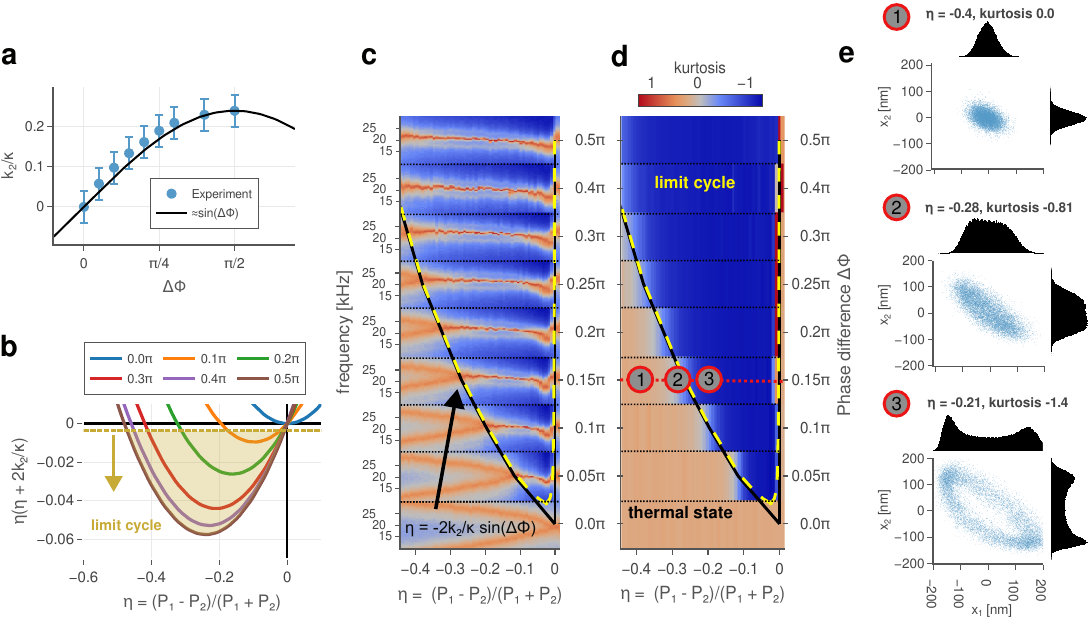}
  \caption{{\bf Tunability of the non-reciprocal interaction force using optical phase difference $\Delta \Phi$.} 
   {\bf a,}  Comparison of the experimental (blue dots) and theoretical (black curve) non-reciprocal coupling rates, $k_2/\kappa$, at different values of the optical phase difference, $\Delta \Phi$. These mean values together with standard deviation (the errobars) were determined using least squares method. 
  {\bf b,} Theoretical prediction of the parametric space where the limit cycle emerges. (curve color denotes the phase difference $\Delta \Phi$)
  {\bf c,} The sum of the power spectral densities (PSD) of $x$-coordinates of the nanoparticles for different values of power detuning, $\eta$, and several values of the phase difference $\Delta \Phi$ illustrating the tunability of the non-reciprocal interaction. Black curves correspond to the exceptional points $\eta = 0$ and $-2k_2/\kappa$ for given $\Delta \Phi$. Dashed yellow curves show the theoretical prediction of parameters where limit cycle appears (taken from part b).   {\bf d,}  Color map encoding the value of excess kurtosis of the probability density distribution of the $x$-coordinates of the particles, quantifying departures from statistical normality and identifying a region where the limit cycle experimentally emerged. 
   {\bf e,}  Two-dimensional scatter plots  of the $x$-coordinates, $x_1$ and $x_2$, together with projected 1D histograms illustrating the spatial extent of the motion for three values of the power detuning (close to the exceptional point) for phase difference $\Delta \Phi =0.15$. (Denoted by the red dashed line and circles in part d.) }\label{fig3}
\end{figure*} 

Now, when $\eta=0$ or $\eta=-2k_2/\kappa$, the eigenvalues are purely real and degenerate with $\lambda_{1,2}=\kappa$. Between these EPs, they form a complex conjugate pair, $\lambda_1=\lambda_2^\ast$, and outside this range, they are real and distinct. Similarly, the oscillation eigen-frequencies, Eq. (\ref{eq:omi0}), are either real and distinct, or form complex conjugate pairs. In the latter case, the imaginary parts of the $\omega_i$ represent gain or loss, depending on the sign. Positive values of $\Im(\omega_i)$ relate to free oscillations in which the motion of the non-conservative force works against the motion of the particles (see dashed curve in Fig.~\ref{fig1}c), reducing their kinetic energy so that they spiral into the fixed point at $x_1=x_2=0$. For $\Im(\omega_i)<0$, the force drives the particles which increase in kinetic energy and spiral outwards (Supplementary Note 6). These processes should be contrasted with the classical PT symmetric oscillators discussed by Bender et al. \cite{bender_systems_2014}, for which the gain and loss mechanisms derive from positive or negative dissipation (i.e. drag). This transitional behaviour is experimentally explored in Figs.~\ref{fig1}b-c. Figure~\ref{fig1}b, shows the sum of the power spectral densities (PSDs) of the particles, revealing the PT-like transition itself, bounded by the exceptional points (EPs) at $\eta=0$ and $\eta=-2k_2/\kappa$. By measuring the range of $\eta$ between the EPs, and comparing with the oscillation frequencies, we can estimate the coupling constant, $k_2$, and the trap stiffness, $\kappa$ (Supplementary Note 5).
Figure~\ref{fig1}c shows the effective damping in the system. The continuous and dashed lines show values of $\Im(\omega_i)$ evaluated from the measured values of $k_2$, $\kappa$ and the detuning $\eta$, Eq. (\ref{eq:lambda}). The experimental data points show a second estimate of the effective damping, taken from the decay rate of the auto-correlation, see Fig.~\ref{fig1}c. (Supplementary Note 5). 
Figures~\ref{fig1}d,e demonstrate the stabilizing effect of the viscous drag $\xi$. When $\xi$ is included in Eq. (\ref{eq:Langf0}), the oscillation eigen-frequencies are approximately,
\begin{equation}\label{eq:omi1}
\omega_i \approx \pm\sqrt{\frac{P_\mathrm{tot}}{2m}} \Re(\sqrt{\lambda_{1,2}}) + \mathrm{i}\Big(\pm\sqrt{\frac{P_\mathrm{tot}}{2m}}\Im(\sqrt{\lambda_{1,2}})+\frac{\xi}{2m} \Big),
\end{equation}
see Supplementary Note 4. Now, all of the $\Im(\omega_i)$ can remain positive, so that the fixed point remains stable, even when the eigenvalues of the stiffness are complex conjugates. The stability condition with finite viscosity can be written as
\begin{equation}
\eta(\eta+2k_2/\kappa)<-\frac{2\xi^2}{mP_\mathrm{tot}\kappa}.
\end{equation}

This condition is shown graphically in Fig.~\ref{fig1}d.  For higher pressures, where $\xi$ exceeds the threshold value, the coordinates of the particles fluctuate around the stable fixed point, $x_1=x_2=0$. As the pressure is lowered, there is an abrupt change. The viscous drag becomes insufficient to compensate for the work done on the particles by the non-conservative optical force, and inertial forces grow until they exceed the confining optical potential. At this point the fixed point becomes unstable and the system undergoes a Hopf bifurcation \cite{simpson_stochastic_2021}. The amplitudes of the oscillations grow until non-linearities in the forces permit the formation of  stable, self-sustained oscillations, or limit cycles, which combine motions of both particles, see Fig.~\ref{fig1}e. The formation and stability of these limit cycles is discussed in Supplementary Note 8.

After the Hopf-bifurcation the system exhibits some of the statistical properties characteristic of phonon lasers \cite{kuang_nonlinear_2023,pettit_optical_2019,zheng_arbitrary_2023,vahala_phonon_2009, sharma_mathcalpt_2022}, i.e. it has a threshold condition for lasing, a narrowing line width and coherent oscillations quantified by the value of the second-order autocorrelation function at zero time delay.
We note that the physical mechanisms behind the statistics are fundamentally different. While the phonon laser is excited by a negative dissipation, proportional to velocity, our system is driven by non-conservative Newtonian forces. Figure~\ref{fig2} illustrates the behaviour of our system above threshold, after the Hopf bifurcation (at pressure 5\,mbar), following the formation of the noisy limit cycles. In this regime, the oscillation frequencies (shown in the summed PSDs in Fig.~\ref{fig2}a) are slightly modified due to the non-linearities in the force that become significant as the oscillation amplitude grows, but retain the topological features they had at higher pressure (Fig.~\ref{fig1}b). These limit cycles consist of closed loops in a four dimensional phase space, $(x_1,x_2,v_1, v_2)$. Figures~\ref{fig2}b,c show sections through this phase space, as the power detuning, $\eta$, is varied. Although it is challenging to directly measure the optical forces, we can measure the work that they do over the course of a limit cycle. For stability, the work done by the non-conservative optical forces must be balanced, on average, by the energy dissipated into the surrounding gas, i.e.
\begin{equation}
W_{d}(t)=\int^t_0 \xi |\vvec|^2 \mathrm{d}t.
\end{equation}
Figure~\ref{fig2}d shows the distribution of this work, $W_{d}(T)$, calculated from the trajectories of the particles over one time period $T$.  For an ideal system, with perfectly non-reciprocal forces, the energy dissipated by the first particle is the same as that dissipated by the second (Supplementary Note 7). However, small non-idealities in our system result in differential dissipation rates, with the non-conservative force performing more work in one trap than the other. In general, both the dissipated energies and their respective variances increase as $\eta$ approaches the second EP at $\eta\approx 0$.


In Figures \ref{fig2}e-g, we probe the statistics of these noisy limit cycle oscillations with several revealing metrics. 
Figure~\ref{fig2}e, shows the relative phase of the oscillations of the particles, which varies from $-\pi$ to $0$, across the region of degeneracy, as predicted by the theoretical model (Supplementary Note 3). We also plot the relative Shannon entropy~\cite{tass_detection_1998, brzobohaty_synchronization_2023}. In this context, $S_r$ measures the strength of phase locking between the coupled oscillations, taking values between zero and one, where a value of one indicates perfect locking.  Figures \ref{fig2}f,g focus on an analogy between our system and the phonon laser. In Figure \ref{fig2}f, we show the line width of the PSD, decreasing significantly from kHz to Hz. We also show the phonon population, $\langle n \rangle = \langle \varepsilon \rangle/(\hbar\Omega) =  1/(\hbar\Omega)\sum_i \varepsilon_i p(\varepsilon_i) \Delta \varepsilon $, where $p(\varepsilon_i)$ is probability density function calculated from the total energy of the particle, i.e. $\varepsilon = (1/2)m \Omega^2 (x^2 + (v/\Omega)^2)$. The oscillatory frequency $\Omega$ was determined from the fit of autocorrelation function calculated  from particle trajectory. 
The phonon population mainly carries information about the increase of the limit cycle amplitude. We note that this parameter varies gradually across the region of degeneracy, but shows an abrupt transition at the second EP (at $\eta=0$). In the terminology of phonon lasers the second-order auto-correlation $g^{(2)}(0) = ( \langle n^2 \rangle - \langle n \rangle)/\langle n \rangle^2$ is used to characterize the transition from a thermal state ($g^{(2)}(0) = 2$) to a coherent one ($g^{(2)}(0) = 1$). In our system we observed a similar transition from thermal state (out of the degeneracy region) to coherent state in the degeneracy region where the limit cycle is stable, see Fig.~\ref{fig2}g. Even though the underlying physical mechanisms are fundamentally different, we observe a striking similarity between the stochastic dynamics observed in our system, and that measured for phonon lasers  \cite{kuang_nonlinear_2023,pettit_optical_2019,zheng_arbitrary_2023}. 

In Figure~\ref{fig3} we demonstrate the extraordinary tunability of our system. 
We vary the strength of the non-reciprocal coupling.  
In Fig.~\ref{fig3}a, we plot $k_2/\kappa$ against $\Delta \Phi$, revealing a sinusoidal variation, consistent with that predicted by the simple dipole model, Supplementary Note 2. 
For comparison, Fig.~\ref{fig3}b shows the theoretical stability condition for varying $\Delta \Phi$. 
Figure~\ref{fig3}c shows the summed PSDs for various values of $\Delta \Phi$ between 0 and $\pi/2$ together with exceptional point (full black curve) and stability region (yellow dashed curve). 
Figure~\ref{fig3}d  shows the value of the excess kurtosis, which has zero value for thermal state with a normal (Gaussian) distribution and negative values about -1 when limit cycles are formed.
The limit cycle is formed for theoretically predicted parameters (see yellow dashed line). 
Figure~\ref{fig3}e compares the probability density function of particle position ($x_1$ and $x_2$) for three values of power detuning close to the exceptional point (see red circles in Fig.~\ref{fig3}d), showing an abrupt change of particle position distribution from the thermal state with Gaussian shape to limit cycle with the non-Gaussian shape together with value of excess kurtosis sharply dropped from 0 to $<-1$ (phase difference $\Delta \Phi = 0.15$).


\section{Conclusion}\label{sec6}
In conclusion, we present a flexible, scalable and reconfigurable optomechanical system capable of supporting the rich dynamical effects associated with the coupled oscillations of multiple mesoscopic particles having controllable level of coupling that range from purely conservative to purely non-reciprocal. 

In the absence of viscous damping, and the accompanying thermal fluctuations, our system exhibits a trivial PT-symmetry. In the presence of weak damping, this symmetry is broken, but the key characteristics of PT symmetric systems are retained. In particular, variations in power detuning take the system from an equilibrium phase (connected with real and distinct eigen-values) through a non-equilibrium phase (connected with complex conjugate eigenvalues), before returning to an equilibrium phase. It is revealing to compare our system with a truly  PT symmetric system of coupled classical oscillators, e.g. \cite{bender_systems_2014,bender_observation_2013}. For the true PT system, the required symmetry is imposed by a balance of positive and negative dissipation, and the associated phase transition occurs when the coupling between the oscillators is varied. In our system, gain and loss mechanisms arise due to the ways in which different oscillations interact with the non-conservative force: those that move with the force, or are driven by the force, are amplified while those that run counter to the force are damped. By contrast to the true PT system, the transition we observe is induced by tuning the resonant frequencies of the coupled oscillators; the non-reciprocal coupling remains approximately constant. In our system, the non-equilibrium phase is stabilized by the finite viscosity. At higher pressures, it is characterised by biased stochastic motion with fluctuations about a stable fixed point. Reducing the pressure destabilizes the fixed point, precipitating a Hopf bifurcation and the formation of a noisy limit cycle oscillation in which the motions of both particles are combined.

The behaviour of the system is analogous to that of the phonon laser \cite{kuang_nonlinear_2023,pettit_optical_2019,zheng_arbitrary_2023,vahala_phonon_2009, sharma_mathcalpt_2022}, with the statistics of the excited, non-equilibrium state being very similar, despite differences in the underlying physical mechanisms (i.e. the phenomena we observe derive from non-conservative Newtonian forces, rather than positive and negative dissipation, as before). 

Finally, the wide tunability of optical interaction between levitated nanoparticles enables the relatively easy extension to many particle systems by creating additional optical traps and detectors, providing a well-controllable platform for the exploration of higher-order exceptional points with possible sensing applications \cite{hodaei_enhanced_2017}, and the engineering of discrete optomechanical media with topological properties such as the non-Hermitian skin effect \cite{zhang_review_2022, zhang_universal_2022, liang_dynamic_2022, zou_observation_2021}, or even optomechanical time crystals similar to those demonstrated by Zheludev \cite{liu_photonic_2023, raskatla_continuous_2023}, but levitated in vacuum.

{\bfseries Note: }  We are aware of similar findings of Reisenbauer et al. \cite{reisenbauer_non-hermitian_2023} where the interaction of nanoparticles levitated in optical tweezers in vacuum are investigated along  optical axis. 

\section{Acknowledgement}
The Czech Science Foundation (GF21-19245K, O.B.);  Akademie v\v{e}d \v{C}esk\'{e} republiky (Praemium Academiae, P.Z.); 
Ministerstvo \v{S}kolstv\'{i} ml\'{a}de\v{z}e a t\v{e}lov\'{y}chovy ($\mathrm{CZ.02.1.01/0.0/0.0/16\_026/0008460}$).

\section{Author Contributions}
S.H.S., O.B., and P.Z. designed and developed the study from the theoretical and experimental aspects, S.H.S. provided theoretical content,
V.L., T.Z.,  and P.J. upgraded the experimental setup and performed the measurements, S.H.S., O.B., V.L., M.\v{S}. and P.Z. analysed the experimental data and compared them to the theoretical results. S.H.S., O.B., V.L. and P.Z. contributed to the text of the manuscript. 

\section{Competing interests}
The authors declare no competing interests.

\section{Methods}\label{sec4}

\subsection{Experimental details}
\subsubsection{Experimental set-up}
 A collimated Gaussian beam (vacuum wavelength 1550 nm) propagating from a laser (Koheras Adjustik) was expanded by a telescope 
(lenses L1 and L2 of focal lengths $f_1= 100$~mm and $f_2=200$~mm) and projected on a digital micromirror device (DMD, Vialux).

The amplitude mask encoded at the DMD diffracted the beam into the $\pm1$ diffraction orders that were used to generate the two counter-propagating 
trapping beams and to control the phase, distance, and balance of power among optical tweezers.
Through our unique DMD-based optical trapping setup we were able to conduct our experiments with high precision and flexibility.

These beams passed through the aperture placed in the focal plane of 
the lens L3 ($f_3=400$~mm) while the zeroth and higher orders were blocked here. 
\begin{figure}[htb]
	\centering
	\includegraphics[width=1\linewidth]{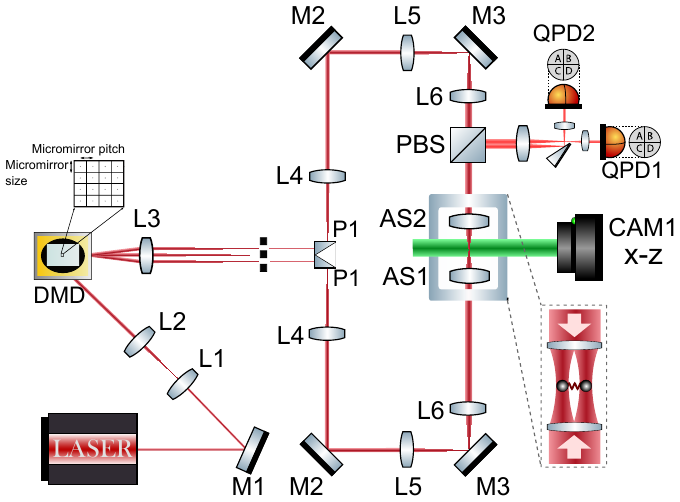}
	\caption{{\bf Experimental set-up} Two pairs of interfering counter-propagating laser beams form standing wave optical traps between the aspherical lenses AS1,2 placed in the vacuum chamber (see the inset in the right bottom). 
    Positions of the nanoparticles in the $x-z$ plane are magnified by an objective and observed by CAM1. Positions of each nanoparticle in the $x-y$ plane are independently 
    but synchronously recorded by quadrant photo-detectors QPD1,2. Digital micro-mirror device (DMD) allows fast modification of laser trapping powers (i.e. setting the power detuning $\eta$), separation $d_0$, and phase difference $\Delta \Phi$ of both pairs of the trapping beams. $\mathrm{L_i}$ and $\mathrm{M_i}$ denote lenses and mirrors forming the trapping beams, PBS is the polarizing beam splitter and $\mathrm{P_1}$ denotes reflecting prisms. 
	}
	\label{Fig:setup}
\end{figure}

The two transmitted
beams were reflected from prisms P1 and collimated by lenses L4 ($f_4=300$~mm). These lenses formed telescopes together with the lens L3 and ensured
that the DMD and mirrors M2 planes become conjugated. Similarly, telescopes consisting of lenses L5  ($f_5=150$~mm) and L6  ($f_6=200$~mm) ensured
conjugation of mirror M2 planes with the back focal planes of aspheric lenses AS1  ($f=8$~mm, maximal NA = 0.5).
We used Thorlabs achromatic doublets AC254-XXX-C (L1 -- L6) and aspheric lenses C240TME-C (AS1) with antireflection coatings and dielectric mirrors BB1-E04 (M1 -- M3).
AS1 focused the beams inside the vacuum chamber and together with the DMD diffraction patterns 
provided the total trapping power and the beam waist radii of $2P=120$\,mW and $w_0 = 1.5\,\mu$m, respectively.


\subsubsection{Particles loading}
Silica particles (Microparticles, mean diameter 611~nm) were dispersed in isopropyl alcohol and sonicated for $\sim30$ min. 
The suspension was loaded onto the ultrasonic nebulizer (Beurer IH 50) and the formed droplets containing the particles were sprayed 
into the trapping region in the vacuum chamber.
By controlling the concentration and flow rate we ensured the regular loading of two particles into the optical traps. 
We initially trapped a single nanoparticle within one optical tweezer, then meticulously adjusted the nebulizer's flow rate to capture 
a second particle in an additional trap, ensuring that this process did not disrupt the positioning of the first nanoparticle. 
When evacuating the vacuum chamber, we switched to cross-polarized beams, repositioning the particles into the center of the overlapping beams. 
Following this, we revert back to a standing wave configuration to ensure the stability of the trapped particles.

\subsubsection{Particles position detection}
Two quadrant photo-diodes (Hamamatsu Photonics, G6849) QPD1,2 coupled with a d-shaped edge mirror were used to record independently but synchronously the motion of the particles in $x-y$ plane. 
This setup allowed us to discern the signal from the two particles separately. 
The QPDs detected light scattered by the trapped particles and generated signals corresponding to the particle positions. 
The d-shaped edge mirror enabled the separation of signals from each particle, thereby reducing cross-talk and providing more accurate measurements.
The sampling frequency was 400~kHz.


In parallel, the particles were illuminated by an independent laser beam (Coherent Prometheus, vacuum wavelength 532~nm, 
beam waist radius $w_0 =50\,\mu$m, power 5~mW at the sample) which enabled imaging and recording the motion of the particles in $x-z$ plane 
by a fast  CMOS camera (Vision Research Phantom V611, the exposure time and frame rate were set 2 $\mu$s and 400 kHz, respectively).
The low power of the illuminating laser ensured negligible contribution to the net optical force acting on the particles.
Our digital microscope was calibrated using a calibrated target, and thus we use it for precise setting of particles distance and by comparing the parallel records from the camera and the QPDs we were able also to transform the QPDs signals from Volts to meter.

The offline tracking of the particle position from the high-speed video recordings was based on the determination of symmetries in the particle 
images~\cite{leite_threedimensional_2018}, which provided us with the information about the in-plane $x$ and $z$ coordinates. This method enabled us to track particles with nanometer precision.\\
We note a subtlety in the experimental identification of attractors, such as limit cycles, in noisy dissipative systems. For deterministic systems, in which trajectories are tracked precisely, the designation is clear and unambiguous. For noisy systems, however, limit cycle oscillations can be hard to distinguish from more complex attractors. For example, quasi-periodic motion has been theoretically predicted in a related system \cite{svak_transverse_2018}. In our case, the identification of limit cycle oscillation rests on the combination of experimental measurement with numerical simulation and stability analysis of the deterministic equations of motion (see Supplementary Note 7). However, quasi-periodic motion remains a possibility when the system is driven with higher optical power.

\subsubsection{Coulomb interaction}
Each particle is randomly charged. To estimate the magnitude of the Coulomb interaction between the particles we performed charge calibration \cite{magrini_real-time_2021, liska_cold_2023}. 
The typical number of elementary charges on the particles was determined to be less than 100  and thus the magnitude of coupling rate $b/\Omega_0 = -q_1q_2/(8\pi\varepsilon_0 m \Omega_0^2d^3)< 10^{-4}$, for Coulomb interaction \cite{rudolph_force-gradient_2022,rieser_tunable_2022} is  3 order of magnitude smaller than that one obtained for optical binding interaction \cite{liska_cold_2023}.

\section{Data availability} 
All data that support the findings of this paper have been deposited to the Zenodo (10.5281/zenodo.11119873) \cite{liska_observations_2024}.

\end{document}